\documentclass[elsart12,epsf,preprintnumbers,eqsecnum]{revtex4}
\usepackage{epsfig,epsf}
\usepackage{verbatim}
\usepackage{amsmath}
\usepackage{amssymb}
\usepackage{graphicx}
\usepackage{enumerate}
\usepackage{slashed, bm}
\usepackage{hyperref, xcolor}

               \let\p = \pi                  
                
\newcommand{\nn}{\nonumber}
\newcommand{\beq}{\begin{equation}}
\newcommand{\eeq}{\end{equation}}
\newcommand{\bea}{\begin{eqnarray}}
\newcommand{\eea}{\end{eqnarray}}

\DeclareMathOperator{\Trace}{\mathnormal{Tr}}

 \newcommand{\eps}{\epsilon} %
 
\newcommand{\xt}{{ \bm x}}
\newcommand{\yt}{{ \bm y}}
\newcommand{\zt}{{ \bm z}}

\newcommand{\qt}{{ \bm q}}
\newcommand{\pt}{{ \bm p}}
\newcommand{\kt}{{\bm k}} 
\newcommand{\lt}{{\bm l}}
\newcommand{\qp}{{q^+}} 
 
\newcommand{\pp}{{p^+}}

\newcommand{\kp}{{ k^+}}

\newcommand{\btil}{\tilde{b}}
\newcommand{\ketv}{\vert 0\rangle}
\newcommand{\brav}{\langle 0 \vert}

\begin{document}

\title{Entanglement entropy, entropy production and time evolution in high energy QCD}
\author{Alex Kovner$^{1,2}$, Michael Lublinsky$^3$, Mirko Serino$^3$}

\affiliation{
$^1$ Physics Department, University of Connecticut, 2152 Hillside Road, Storrs, CT 06269-3046, USA \\
$^2$ CERN, Theoretical Physics Department, CH-12-11 Geneva 23, Switzerland \\
$^3$ Department of Physics, Ben-Gurion University of the Negev,\\
Beer Sheva 84105, Israel\\
} 

\preprint{CERN-TH-2018-131}

\begin{abstract}
Working in the framework of the Color Glass Condensate effective theory of high energy QCD, 
we revisit the momentum space entanglement entropy of the soft gluons produced in high energy dilute-dense  collisions.
We extend  the work of~\cite{Kovner:2015hga} by considering entropy produced in a single event. 
This entropy arises due to decoherence of eigenstates with different energies during the time evolution after the collisions with the target. 
We define it rigorously as the entanglement entropy of the produced system with the experimental apparatus.
We compute the time dependent single event entropy in the limit of weak projectile field. 
Further we compute the entropy for the ensemble of events defined by the McLerran-Venugopalan model for the projectile wave function. 
Interestingly the entropy of the ensemble has a much weaker time dependence than the entropy in any single event. 
We attribute this feature to the so called monogamy of entanglement.
\end{abstract}

\date{\today}

\maketitle

\mbox{}

\pagestyle{plain}
\setcounter{page}{1}

\section{Introduction}

Entanglement is a unique feature of quantum systems, which distinguishes  them profoundly from classical ones.
In a quantum system a convenient measure of entanglement is the entanglement entropy.

In quantum theory one defines the entanglement entropy between two sets of degrees of freedom, say $A$ and $B$, 
which span the Hilbert space of the theory $\mathcal{H}$ in the sense that  $\mathcal H = \mathcal H_A \otimes \mathcal H_B$, 
where $A$ and $B$ span $\mathcal H_A $ and $\mathcal H_B $ respectively. 
The entanglement entropy measures to what extent the two sets of degrees of freedom are entangled in a given state, 
and thus to what degree the state of $A$ becomes mixed upon ``integrating out'' the degrees of freedom $B$.

In the last two decades this topic has attracted a lot of interest  from various physics perspectives~\cite{Calabrese:2004eu,Calabrese:2009qy}.
Recently some ideas have been put forward about possible importance of entanglement  also in the context of high energy scattering processes. 
In particular it has been suggested  that the observed apparent thermal nature of the produced soft particle spectra 
at colliders might be caused by entanglement of the different degrees of freedom in the  hadronic wave function. 
Since the bulk of hadronic degrees of freedom are entangled with those that are probed in the scattering process, 
they may effectively  act as a thermal bath even though no actual thermalization occurs since there is no interaction in the final state ~\cite{Baker:2017wtt,Kharzeev:2017qzs}. 
Related development has been reported in~\cite{Berges:2017zws,Berges:2017hne} where a one dimensional toy model has been studied from a similar point of view. 
A commonality of these approaches is that they are concerned with the entanglement entropy between spatial regions,
as it is also customary in almost all the rest of the field theoretical  literature.
On the other hand, the recent work of~\cite{Hagiwara:2017uaz} has investigated the entropy of parton distribution functions,
particularly their low-$x$ behaviour, a topic already addressed in the early attempt ~\cite{Kutak:2011rb} to compute the entropy of a gluon system 
produced in a high energy proton-proton collision. Similar earlier ideas along these lines were discussed in~\cite{Elze:1994hj,Elze:1994qa,Peschanski:2012cw},
whereas~\cite{Liu:2018gae} investigates entanglement in high energy parton-parton scattering using the $AdS/CFT$ correspondence.

Entanglement entropy in momentum space in relativistic QFT has also been studied, although less extensively. 
Reference ~\cite{Balasubramanian:2011wt} investigated entanglement between modes inhabiting distinct shells of momentum space. 
In the context of high energy scattering entanglement in momentum space is a very natural and interesting property. 
In particular in the framework of the Color Glass Condensate (CGC) approach to high energy evolution, 
the low longitudinal momentum modes are by construction fundamentally entangled with the valence, or high energy modes.  
The crucial feature of the CGC wave function is that the soft gluon part of the wave function is nontrivial due to the presence of the valence color charges. 
The dependence of this wave function on energy is described by the JIMWLK evolution 
equation~\cite{JalilianMarian:1996xn,JalilianMarian:1997jx,JalilianMarian:1997gr,JalilianMarian:1997dw,JalilianMarian:1998cb,Kovner:1999bj,Kovner:2000pt,Weigert:2000gi,
Iancu:2000hn,Iancu:2001ad,Ferreiro:2001qy}.
This entanglement entropy between the soft and valence degrees of freedom was investigated in~\cite{Kovner:2015hga}. 
It was also shown there that this entanglement entropy, albeit not equal, is directly related to the entropy of the particles produced in an energetic hadronic collision. 


In the present paper, we continue the line of investigation initiated in ~\cite{Kovner:2015hga}. 
The main question we are asking is whether it is possible to define sensibly production entropy within the CGC framework on an event-by-event basis.
The calculation performed in ~\cite{Kovner:2015hga} involves averaging over the valence color charge density in the projectile wave function. Since the valence charges are slow degrees of 
freedom (in the sense of slow variation in light cone time), a given event corresponds to a given configuration of the color charges. Thus averaging over the slow degrees of freedom in the projectile 
wave function corresponds to averaging over the event ensemble. 
In this sense the entropy calculated in ~\cite{Kovner:2015hga} should be understood as the entropy corresponding to the whole ensemble of events. 
On the other hand intuitively it is clear that one should be able to ascribe entropy even to a single event. 

Even though in a strict sense the state of soft gluons emerging from a collision in a given event is a pure state, 
it contains a superposition of eigenstates of the theory with very different energies. 
The phases of these eigenstates are not infinitely resolvable by a real experimental apparatus, 
because of its finite resolution power. As a result the information about relative phases of the different eigenstates is scrambled by the measurement.
This implies that the pure state produced by a scattering event will always be known - at best - in terms of a density matrix.
Thus, beside the entanglement of the soft modes with the valence modes investigated in~\cite{Kovner:2015hga}, 
another unavoidable source of incomplete information exists.

 The question is how to describe such a density matrix and the associated entropy  with the tools of quantum mechanics.
Indeed, time is not a quantum mechanical degree of freedom \emph{per se}, which makes it impossible to directly
investigate decoherence in time in terms of a partial trace over a Hilbert subspace - the standard tool in defining a reduced density matrix.

The idea investigated in the present paper relies on the energy-time uncertainty relation.
We study the decoherence in time of the soft gluon state emerging from the scattering event by interpreting it as a consequence 
of an entanglement of the final state particles with an imaginary experimental apparatus. The apparatus has a finite time resolution, which determines the entropy of the given event.
Put differently, we extend the Hilbert space of our system to accommodate one more auxiliary degree of freedom,
 which couples directly to energy.  For simplicity we take the wave function of this ``calorimeter'' degree of freedom to be a gaussian, and will therefore refer to it as a \emph{white noise}.

In this setup we can trace the density operator of the soft gluon system over this auxiliary Hilbert subspace.
This allows us to define an event-by-event entropy of the produced gluons which is induced by their time evolution.
In the sense just specified, this entropy reflects the incomplete information ascribable to the finite time resolution of the experiment $T$.
We analytically compute  the entropy of the soft gluons defined in this way in the limit of weak projectile fields. In the second part of the paper we generalize our discussion to include both the 
event-by-event entropy and the entropy of the ensemble of events discussed in~\cite{Kovner:2015hga}.

The plan of the paper is the following. In section \ref{setstage} we review the basic CGC setup with the McLerran-Venugopalan (MV) 
model as the model for the distribution of valence charges in the projectile wave function.
In section \ref{white} we introduce the main idea of this work, and couple the wave function of the soft gluons emerging after scattering on the target to the ``calorimeter'' or  \emph{white noise} degree of freedom. 
Sections \ref{enenfixedrho} and \ref{enenavgrho} contain the central results of this paper:
we compute the entanglement entropy of the system of produced soft gluons respectively for a fixed 
configuration of the color charge density of the projectile wave function (single event) and for the ensemble of events defined by the MV model~\cite{McLerran:1993ni,McLerran:1993ka}
and averaged over the target fields. Both computations are performed by resorting to the so-called \emph{replica trick}.
We conclude in section \ref{conclusions} with short summary of our results.

\section{The Setup}\label{setstage}
\subsection{The CGC state}

In this paper we work within the framework of the high-energy limit of dilute-dense QCD~\cite{Gelis:2010nm}. 
The fast hadron, referred to as the projectile, is  highly boosted in the $+$ direction and is properly described by a wave function $\psi$ which can be written in a quasi factorised form 
\beq\label{wf}
\vert \psi \rangle = \vert s \rangle \otimes \vert v \rangle \, , 
\eeq
Here $\vert v \rangle$ is the the wave function that depends only on the valence (energetic) degrees of freedom, 
while $ \vert s \rangle$ is the wave function of the soft gluons.  The relevant valence degrees of freedom are described by 
the set of classical color  charge densities $\rho^a(\xt)$ which only depend on the transverse coordinates $\xt$. 

The wave function in eq.(\ref{wf}) is not a product wave function, since the soft part of it depends on the valence charges. 
This entanglement between the energetic and soft modes is precisely the source of the entropy studied in  ~\cite{Kovner:2015hga}.

As long as the color charge density is not too large, a good approximation for the soft part of the wave function 
at LO in the strong coupling constant is given by a simple coherent state~\cite{Kovner:2005jc}
\beq\label{soft}
\Omega \ketv = 
\exp\left\{i\int _{q^+<\Lambda}
\btil^i_a(q)\, \left[a^a_i(\qp,\qt) + a^{\dagger a}_i(\qp,-\qt) \right] \right\}\vert 0\rangle \, ,
\eeq
As opposed to ~\cite{Kovner:2015hga} we will have to follow the longitudinal momentum dependence of various modes. 
We will therefore explicitly keep track of the longitudinal momentum dependence of the classical field
and, thus, we have introduced the longitudinal momentum dependent classical field
\beq
\btil^a_i(q) \equiv \sqrt{\frac{2}{\qp}}\, b^a_i(-\qt)\, ,
\eeq

The standard Weizs\"acker-Williams (WW) field generated by the transverse valence color charges, respectively in coordinate and momentum space, is
\bea
b^a_i(\xt) 
&=&
\frac{g}{2\pi}\, \int d^2\yt \frac{(\xt-\yt)_i}{(\xt-\yt)^2}\, \rho^a(\yt) \, ,
\nn \\ 
b^a_i(\qt)  
&=& 
-\frac{i\,g\,\rho^a(\qt)\, \qt_i}{\qt^2}\, ,
\nn \\
{b^a_i}^\dagger(\qt) 
&=& 
b^a_i(-\qt)\, ,  
\label{WWtrans}
\eea
In the following sometimes we will use the mixed representation of $\btil^a_i$ in the $(q^+, \xt)$ space.

The scale $\Lambda$ in eq.(\ref{soft}) is the longitudinal momentum scale that provides the separation between the valence and soft modes. 
The classical color charge density of the valence modes is as usual
\bea
\rho^a(\qt) 
&=& 
-i f^{abc}\, \int_{\kp>\Lambda} \frac{d\kp}{2\p} \int \frac{d^2\kt}{(2\pi)^2}\, a^{\dagger\, b}_j(\kp,\kt)\, a^c_j(\kp,\kt+\qt) \, ,
\nn \\
{\rho^{a\,\dagger}(\qt) }
&=& 
\rho^a(-\qt)\, .
\eea
This notation is the same as in~\cite{Kovner:2015hga}.

Whenever integrating over the $3$-momenta of the soft modes we use the shorthand notation
\beq
\int_k \equiv \int_{\kp<\Lambda} \frac{dk^+}{2\pi} \int \frac{d^2\kt}{(2\pi)^2} \, , 
\eeq
%

%
%

The creation and annihilation operators satisfy the fundamental commutation relation
\beq
[ a^a_i(\qp,\qt) ,a^{\dagger b}_j(\pp,\pt) ] = (2\pi)^3 \,\delta^{ab} \, \delta_{ij}\,  \delta(\qp-\pp) \, \delta^{(2)}(\qt-\pt)\, .
\label{basicomm}
\eeq
%
%
%
%
%
We also introduce the following usual combinations of creation and annihilation operators,
\bea
\phi^a_i(\qp,\qt) 
&\equiv&
a^a_i(\qp,\qt) + a^{\dagger a}_i(\qp,-\qt) \, ,
\nn \\
\pi^a_i(\qp,\qt)  
&\equiv& 
-i \left( a^a_i(\qp,\qt)\right) - a^{\dagger a}_i(\qp,-\qt) \, ,
\nn \\
\left[ \phi^a_i(\qp,\qt), \pi^b_j(\pp,\pt) \right]   
&=&
2\,i\, (2\pi)^3 \delta^{ab}\,  \delta_{ij}  \, \delta(\qp-\pp)\, \delta^{(2)}(\qt+\pt) \, , 
\label{phidef}
\eea
Since the operators $\phi^a_i(\qp,\qt) $ form a mutually commuting set we will choose throughout this paper the basis of their eigenstates as the basis in which to write the density matrix. The free vacuum state $\vert 0\rangle$ entering eq.(\ref{soft}) in this basis has the form
\beq
\langle \phi |0 \rangle \equiv \psi(\phi) = 
\mathcal{N}\, \exp\left\{- \frac{1}{4}\int_{\qp\qt} \phi^a_i(\qp,\qt) \phi^a_i(\qp,-\qt)\right\}  \, .
\label{vwfmom}
\eeq
%
%

Note that the coherent operator $\Omega$ appearing in eq.(\ref{soft}) diagonalizes the light cone Hamiltonian of QCD in the first order in perturbation theory. 
It is also equal to the (appropriately regularized) Heisenberg picture time evolution operator of QCD. 
These points are reviewed in Appendix A.

So far we have discussed the structure of the soft gluon state. To calculate physical observables clearly one also needs information about the state $\vert v\rangle$. This information is non perturbative and therefore has to be supplied independently of the perturbation theory used to calculate the soft wave function eq.( \ref{soft}).
In this paper, just like in~\cite{Kovner:2015hga}, we will use the MV model. 
This model does not specify the valence wave function completely, but it is sufficient for our purposes since it does specify the diagonal matrix elements of the valence space density matrix. 
In the MV model one treats the color charge density as classical, and the value of the color charge density field specifies the basis of states in the valence part of the Hilbert space. 
The MV model then postulates
\begin{equation}
\langle\rho^a(\xt)\vert v\rangle\langle v\vert\rho^a(\xt)\rangle=N\exp\left\{-\frac{1}{2}\int d^2\xt\, d^2\yt\, \rho^a(\xt)\mu^{-2}(\xt-\yt)\rho^a(\yt)\right\}\, ,
\end{equation}
with the normalization constant $N$ such that $\langle v\vert v\rangle=1$.

\subsection{The eikonal scattering}

When scattering on a dense target, the partons of the projectile undergo eikonal scattering
\begin{equation}
a^a(\xt,p^+)\rightarrow S^{ab}(\xt)a^b(\xt,p^+);\ \ \ \ a^{\dagger a}(\xt,p^+)\rightarrow S^{ab}(\xt)a^{\dagger b}(\xt,p^+); \ \ \ \rho^a(\xt)\rightarrow S^{ab}(\xt)\rho^b(\xt) \, .
\end{equation}
The eikonally scattered CGC soft ground state therefore is
\begin{equation}\label{wf1}
\hat S\, \Omega\, \vert 0\rangle\otimes\vert v\rangle=\exp\left\{i\int _{q^+<\Lambda}\int_{\bf x}\btil^{'a}_i(q^+,{\bf x})\, S^{ab}(x)\phi^b(q^+,{\bf x})\right\}\vert 0\rangle  \otimes \hat S\vert v\rangle\, ,
\eeq
where the Weizs\"acker-Williams field of the eikonally rotated charge is
\beq
b^{' a}_i(\xt) = \frac{g}{2\pi}\, \int d^2 \yt \frac{(\xt-\yt)_i}{(\xt-\yt)^2}\, \bar\rho^a(\yt) \, , 
\label{scatterWW}
\eeq
where
\beq
\bar \rho^a(\yt)= S^{ab}(\yt)\, \rho^a(\yt) \, .
\eeq
Note that this is the wave function which emerges immediately after the scattering with the target. 
Between the scattering time and the observation time it evolves with the ``free'' QCD Hamiltonian, so that at any time $t$  we have
\beq\label{psi}
\Psi_{out}=U(0,t)\, \hat S\, \Omega\, \vert 0\rangle\otimes \vert v\rangle; \ \ \ \ U(0,t)=\exp\{-iHt\} \, ,
\eeq
where $H$ is the  QCD light cone Hamiltonian, defined in Appendix \ref{coherent} (here and throughout this paper we denote the light cone time variable by  $t$).
Also the relation of $\Omega$ to the time evolution operator $U(t,0)$is clarified in Appendix \ref{coherent}.

\subsection{The soft gluon density matrix}

In this paper we are interested in the density matrix of soft gluons. 
Given the wave function eq.(\ref{psi}), the density matrix describing the ``remnants'' of the projectile following the high energy collision can be written as
\beq\label{rho}
\hat{\rho}_P =  U(0,t)\, \hat S\, \Omega \ketv \otimes\vert v\rangle\langle v\vert\otimes\brav \Omega^{\dagger}\, \hat S^\dagger\, U^{\dagger}(0,t)\, 
\eeq

This however is not quite the density matrix that is of interest to us. We are interested in the density matrix that describes the distribution  the soft gluons produced in the scattering.
In particular in the second part of this paper we will be interested in the reduced density matrix obtained after integration over the valence degrees of freedom. 
Naively, one would think that all one has to do is to trace $\hat\rho_P$ over the color charge density. 
However this is not quite right. Recall that the valence color charge density is always accompanied by its ``native'' WW soft field. 
This WW field does not describe soft gluons produced in the collision, but is a part of the  wave function of receding energetic fragments of the projectile. 
We are not interested in this part of the wave function. In other words what we need to trace over is not just the color charge density, 
but the color charge density dressed by its WW field. As discussed in~\cite{Kovner:2015hga}, 
the change of basis from the color charge density to the color charge density accompanied by its WW field is affected by the unitary operator $\Omega$. 
Therefore our object of interest is not the density matrix of eq.(\ref{rho}), but rather a unitarily transformed matrix

\beq
\hat{\rho}_{P}'= \Omega^\dagger \, \hat\rho_P \, \Omega = 
\Omega^{\dagger}\, U(0,t)\, \hat S\, \Omega \ketv \otimes\vert v\rangle\langle v\vert\otimes \brav \Omega^{\dagger}\, \hat S^\dagger\, U^{\dagger}(0,t)\, \Omega\, .%
\label{density1}
\eeq
To reiterate, this expression has a simple physical interpretation~\cite{Kovner:2006wr}: the coherent state $\Omega \vert 0 \rangle$
is scattered and then evolved up to time $t$. The inverse of the coherent operator, $\Omega^{\dagger}$, 
associates to the scattered valence modes  their ``native''  gluon cloud, thus subtracting those bound state gluons from the final state.
The resulting soft wave function contains only the gluons which are produced by the scattering process and are not a part of the receding energetic remnants of the projectile. 

We can write this expression in a more explicit form. First, recall that the operator $\Omega$ perturbatively diagonalizes the light cone Hamiltonian (see Appendix A):
\beq
\Omega^\dagger U(0,t) \Omega = \exp\left\{-i\, [H_0 + F(\rho)] t \right\}\, , 
\eeq
where $H_0$ is the free noninteracting Hamiltonian of the soft modes, and the vacuum energy $F(\rho)$ ( defined in (\ref{OmegaH}) ) 
does not depend on the soft degrees of freedom. 
The vacuum energy only adds a pure phase to the soft wave function, whuch we will disregard in the following.
On the other hand using eq.(\ref{wf1}) and the explicit form of the operator $\Omega$ we can write
\beq
\Omega^\dagger\, \hat S\, \Omega\,  \ketv \otimes\vert v\rangle = 
\exp\bigg\{ i\,\frac{g}{2\pi}\, \int d^2\xt\,\int d^2\zt\, \frac{(\xt-\zt)_i}{(\xt-\zt)^2} \, \bigg( S^{ba}(\xt) -S^{ba}(\zt) \bigg)\, 
\bar\rho^b(\zt)\, \int_{q^+}\phi^a_i(q^+,\xt) \bigg\}\, \vert \, 0 \rangle\otimes \hat S\vert v\rangle \, .
\label{deltabintro}
\eeq
Note that
\beq
\langle\rho^a(\xt)\vert v\rangle=\langle \bar\rho^a(\xt)\hat S\vert v\rangle\, .
\eeq
In what comes below, we will change the basis for the valence degrees: $\hat S \vert v \rangle \rightarrow \vert v \rangle$
and simultaneously rename $\bar\rho \rightarrow \rho$. Note that in section \ref{enenavgrho} we will integrate over the distribution
of $\rho$. The MV weight used for this integration is invariant under this change of variables.

Now, introducing
\bea
\Delta b^a_i(\xt)
&\equiv&
\frac{g}{2\pi}\, \int d^2\zt\, \frac{(\xt-\zt)_i}{(\xt-\zt)^2} \, \bigg( S^{ab}(\xt) -S^{ab}(\zt) \bigg)\,\rho^b(\zt) \, , 
\nn \\
\Delta b^a_i(\qt)
&=&
i\,g\,\int \frac{d^2\lt}{(2\pi)^2}\, \bigg[\frac{\qt_i}{\qt^2}-\frac{\lt_i}{\lt^2}\bigg]\, S^{ab}(\qt-\lt)\, \rho^b(\lt)\, ,
\label{deltabform}
\eea

we obtain
\beq
\Omega^\dagger\, \hat S\, \Omega\,  \ketv \otimes \vert v\rangle=  \exp\bigg( i\, \int_{q^+}\int d^2\xt\,\Delta \tilde b^a_i(q^+,\xt)\, \phi^a_i(q^+,\xt) \bigg)  \, \vert \, 0 \rangle \otimes\vert v\rangle\, ,
\eeq

where, as before, $\Delta \btil^a_i(q) = \sqrt{2/\qp}\, \Delta b^a_i(-\qt)$.

Thus, when all said and done, the density matrix that we will be analysing has the form
\beq
\hat\rho_{P}'=e^{-iH_0t} e^{ i\, \int_{q^+}\int d^2\xt\,\Delta \tilde b^a_i(q^+,\xt)\, \phi^a_i(q^+,\xt)  } \, \vert \, 0 \rangle \otimes\vert v\rangle\, \langle v\vert\otimes\langle 0\vert 
e^{- i\, \int_{q^+}\int d^2\xt\,\Delta \tilde b^a_i(q^+,\xt)\, \phi^a_i(q^+,\xt)  }e^{iH_0t}\, .
\label{notnewrho}
\eeq

At $t=0$, this is the density matrix of~\cite{Altinoluk:2015uaa} used to address two-particle correlations and also the starting point of~\cite{Kovner:2015hga}
The focus of~\cite{Kovner:2015hga} was to calculate the entanglement entropy between the valence and soft modes by reducing the density matrix over the valence Hilbert space.
In this paper instead our aim is to define the entropy of soft gluons produced in a single event, as discussed in the introduction. What we mean by a single event, is a fixed configuration of the color charge density $\rho$ and the configuration of the target $S$. 

\section{Density matrix for a single event: the energy resolution and the calorimeter ``white noise''}\label{white}
Formally speaking  eq.(\ref{notnewrho}) at fixed value of $\rho$ and $S$ is a pure state and therefore has vanishing von Neuman entropy. 
However this pure state is a superposition of many states with very different energies. 
When written in the energy basis therefore some of its off diagonal matrix elements are strongly  oscillating functions of time. 
Quite generally, suppose  our system is in a pure state described by a wave function 
\beq
\vert\psi(t)\rangle =
\sum_{n} e^{-iE_n t } c_n \vert \psi_n \rangle \, ,
\eeq
where  $\left\{E_n\right\}$ are energy eigenvalues which, for simplicity, we assume to be non degenerate.
The density matrix of this state in the energy eigenbasis has the form
\beq
\hat\rho(t) =
\vert\psi(t)\rangle\langle\psi(t)\vert = 
\left(
\begin{array}{ccc}
\vert c_1 \vert^2 & c_1 c_2^* \, e^{i( E_1-E_2) t} & \dots \\
& & \\
c_2 c_1^* \, e^{i(E_2-E_1)t} & \vert c_2 \vert^2 & \dots \\
\dots & \dots& \dots \\
\end{array}
\right) \, .
\label{rhonoreduc}
\eeq
If one performs a measurements on this system which takes time $T$ with $T\gg \vert E_1-E_2 \vert^{-1}$, 
such measurement effectively is sensitive only to the time average of the density matrix over $T$. 
Thus all the off diagonal matrix elements between states with large enough energy differences 
effectively vanish for the purpose of such measurement, and our density matrix is equivalent to
\beq
\hat\rho \sim
\left(
\begin{array}{ccc}
\vert c_1 \vert^2 & 0 & \dots \\
& & \\
0 & \vert c_2 \vert^2 & \dots \\
\dots & \dots& \dots \\
\end{array}
\right) \, ,
\label{rhoreduc}
\eeq

Therefore for the purpose of many measurements the pure state of this type  is practically equivalent to a mixed density matrix with vanishing off diagonal matrix elements between eigenstates with 
vastly different energy, where the relevant energy scale is introduced by the time resolution of the measuring apparatus. Such a mixed density matrix has for example  a non vanishing von Neuman 
entropy. In the context of the present paper the wave function $\psi(t)$ is the wave function of a stream of particles emerging from a scattering event which took place at $t=0$. 

Having a finite time resolution is not only inevitable practically, but is required by the Heisenberg uncertainty relation for an experiment that measures particle energies.
Ideally, in the case of an experiment able to resolve energies with great precision, 
the energy-time uncertainty relationship dictates that the time resolution $T$ of the detector should be very big. 
This implies that the off diagonal terms in this "measurement-averaged" density matrix will become practically negligible.

The technical question one should ask is how to sensibly define the ``time averaged'' density matrix for a single event. Directly averaging over $T$ the operator $\hat \rho$ does not seem to be the 
right thing to do, since in general such averaging procedure does not preserve  all the physical properties of the density matrix, such as positivity of its eigenvalues. The preceding qualitative 
discussion suggests however a physical way of doing this. Since the  coherence of the outgoing quantum state is broken due to the interaction with the measuring apparatus, one should introduce 
such an apparatus as a separate degree of freedom coupled to the soft gluons. In particular, if the apparatus is a ``calorimeter'', i.e. measures particle energies, 
then the additional degree of freedom should directly couple to the energy of the state. 
The state of the ``calorimeter degree of freedom'' itself should contain information about the time resolution via some typical time scale. 
We can then define a proper time averaged density matrix of soft gluons by  reducing it over the calorimeter degree of freedom. 
This procedure produces a density matrix that satisfies the correct probability properties and which also incorporates the physics of time averaging of the original $\hat \rho$.

The procedure we have described is not unique, as in principle it depends on the exact state of the calorimeter degree of freedom. However we expect that for all practical purposes it should not matter much how exactly we specify this state as long as it incorporates the appropriate time resolution scale. In the following we implement these ideas by choosing the simplest possible option - endowing the calorimeter degree of freedom with a simple Gaussian wave function. This  corresponds to coupling of the soft gluons to white noise.

The extended density matrix (\ref{notnewrho}) that is now defined on a larger Hilbert space which contains an additional degree of freedom $\xi$ is
\beq\label{rhoxi}
\hat{\rho}_{P,\xi} =  e^{-i H\xi}U(0,t)\, \hat S\, \Omega \vert G\rangle\otimes\ketv \otimes\vert v\rangle\langle v\vert\otimes\brav \otimes\langle G\vert \Omega^{\dagger}\, 
\hat S^\dagger\, U^{\dagger}(0,t) e^{i H\xi}\, 
\eeq
with
\beq
\langle \xi\vert G\rangle=e^{- \frac{\xi^2 }{2T^2}} 
\eeq
As before, dressing this density matrix with the operator $\Omega^\dagger$, and concentrating on the matrix element in the $\xi$-space we are lead to consider
\bea
\langle \xi_1\vert\hat \rho'_{P,\xi}\vert \xi_2\rangle
&=& 
\frac{1}{\sqrt \pi\, T}\, e^{- \frac{\xi_1^2 + \xi_2^2}{2T^2}} 
e^{-i H_0 \xi_1}\, e^{i \int_q \Delta \btil^a_i(q)\, \phi^a_i(\qp,\qt)} \ketv
\brav e^{-i \int_\pt \Delta \btil^b_j(p)\, \phi^b_j(\pp,\pt)} \, e^{i H_0 \xi_2} \, ,
\label{newrho}
\eea
where the normalization is dictated by the requirement $\Trace \hat\rho = 1$. The reduced density matrix of the soft gluons is then calculated as
\bea
 \hat{\rho}'_{P,\xi}
&=& \int_\xi
\frac{1}{\sqrt \pi\, T}\, e^{- \frac{\xi^2 }{T^2}} 
e^{-i H_0 \xi}\, e^{i \int_q \Delta \btil^a_i(q)\, \phi^a_i(\qp,\qt)} \ketv
\brav e^{-i \int_\pt \Delta \btil^b_j(p)\, \phi^b_j(\pp,\pt)} \, e^{i H_0 \xi} \, .
\label{newrho1}
\eea

We stress again that the formal extension of the Hilbert space allows us to interpret the incomplete experimental knowledge of the wave function 
in terms of entanglement between the system and the finite-resolution experimental apparatus.
The effect of this apparatus is to decohere energy eigenstates which have large enough energy differences. 
The decoherence occurs over the time $T$, and thus the $T$-dependence of the density matrix  eq.(\ref{newrho1}) 
can be interpreted in terms of time evolution of the density matrix associated with the final state wave function of the hadronic system in a single event. 
This density matrix has an associated von Neuman entanglement entropy.
This entropy describes the increasing loss of information which is ascribable to the decoherence of different energy modes, in the sense discussed above.

\section{Event by event entropy production in the weak field limit}\label{enenfixedrho}
Our goal now is to calculate the von Neuman entropy associated with the density matrix eq.(\ref{newrho1}). 
The canonical expression of the von Neumann entropy for a system described by a quantum density matrix $\hat \rho$ is
\beq
\sigma^E = - \Trace [\hat\rho \log\hat\rho]\, .
\eeq
This is frequently calculated using the so called replica trick, by representing it as
\beq
- \hat \rho \log \hat \rho = - \lim_{\eps \rightarrow 0} \frac{\hat \rho^{1+\eps} - \hat \rho}{\eps} \, .
\label{replica}
\eeq
and therefore
\beq\sigma^E=-\lim_{\eps \rightarrow 0}\frac{\Trace[\hat \rho^{1+\eps}]-1}{\eps} \, .
\label{repent}
\eeq

%
%
The $\xi$ integral in eq.(\ref{newrho1}) is Gaussian, and therefore can be in principle performed  exactly. This however leads to a complicated integration over the fields $\phi$, and we were not able to calculate the entropy for arbitrary $\rho$. However it turns out to be possible to perform the calculation for small number of produced particles, i.e at small $\Delta\btil^2$. This calculation is the subject of the preset section. Even this calculation turns out to be not completely trivial, as a naive expansion in the small parameter $\Delta\btil^2$ is divergent.

\subsection{Naive replica trick does not work}
At small value of the field $b$ it would seem naively that one can expand the integrand in eq.(\ref{newrho1}) in powers of $\Delta\btil^2$. A little thought however shows that this is not possible.
The issue is that for vanishing $\Delta \btil^2$, $\hat\rho$ is the density matrix of a pure state. 
It therefore has one eigenvalue equal to $1$, while all the other eigenvalues vanish. 
At small but finite  $\Delta \btil^2$ one expects all eigenvalues to change  by an amount proportional to $\Delta \btil^2$.  Thus we expect the density matrix eq.(\ref{newrho1}) 
to have one eigenvalue close to unity $\Delta_1\equiv 1-\delta_1$, and all other eigenvalues $\delta_i$ small, 
so that all $\delta_i\sim \Delta \btil^2$. If this is the case, the main contribution to the entropy arises from the small eigenvalues and it should behave as
\beq
\sigma^E\sim -\Delta \btil^2\ln[\Delta \btil^2]\, .
\eeq
Such an expression clearly cannot be expanded in powers of $\Delta \btil^2$, and thus we expect to hit a divergence if we try to do so. This is indeed what happens. 
We have performed this calculation explicitly and have verified that in the limit $\epsilon\rightarrow 0$ 
the entropy defined in eq.(\ref{repent}) with $\hat\rho$ expanded in $\Delta\btil^2$ diverges.

We will employ a different strategy  than just expanding the entropy in powers of the small parameter. Our logic is the following.
Since corrections to all eigenvalues of $\hat\rho$ are governed by the same small parameter, we expect 
\beq
\delta_{i\geq 2} = \beta_i \, \delta_1 \, , \quad \,  \beta_i\geq 0\, ,\quad\,  \sum_{i\geq 2} \beta_i = 1\, .
\eeq
The second equality follows from the fact that
\beq
\Trace \hat \rho=1-\delta_1+\sum_{i\geq 2}\delta_i=1 \, .
\eeq
In terms of these eigenvalues the entropy is expressed as
\beq
\sigma^E=-\left[(1-\delta_1)\log(1-\delta_1)+\sum_{i\geq 2}\delta_i\log\delta_i\right]= 
-\delta_1 \log \delta_1 - \delta_1 \sum_{i\geq 2} \beta_i \ln \beta_i -(1-\delta_1)\log(1-\delta_1)\approx -\delta_1 \log \delta_1\, .
\eeq
The last equality is valid to leading logarithmic order in the small parameter at hand $\Delta\btil^2$.

Our goal therefore should be to find $\delta_1$. We will use the replica trick to do it, but we will employ the replicas in a somewhat different way than outlined in the beginning of this section.
We note that, since $\Delta_1$ is the largest eigenvalue of the density matrix and the only one that does not vanish at small $\Delta\btil^2$, it must be true that
\beq
\Trace[\hat\rho^N]=\Delta^N_1+O((\Delta\btil^2)^N)\rightarrow_{N\rightarrow\infty}=\Delta^N_1=1-N\delta_1+\frac{N(N-1)}{2}\delta_1^2+...
\eeq
Thus calculating  $\Trace[\hat\rho^N]$ at large $N$, and subsequently expanding in $\Delta\btil^2$, we extract directly $\delta_1$ and therefore the entropy to leading logarithmic accuracy. This will be our strategy.

\subsection{Calculating the entropy}

The first thing to do is to reduce the integrand of this expression to a function of $\phi$, i.e. to get rid of $H_0$ in the exponential. For this we use
\beq\label{phit}
e^{-i H_0 \xi} \phi^a_i(\qp,\qt)e^{i H_0 \xi}= \phi^a_i(\qp,\qt)\cos(E_q\xi)- \pi^a_i(\qp,\qt)\sin(E_q\xi) \, ,
\eeq
with $E_q = \qt^2/2\qp$.
Calculating the matrix element of the density matrix (\ref{newrho1}) we find
\bea
&&\langle\phi^1\vert\hat\rho'_{P,\xi}\vert\phi^2\rangle=
\frac{1}{\sqrt \pi\, T} \int d\xi\, e^{-\int_\qt \frac{\rho^a(\qt)\rho^a(-\qt)}{2\mu^2(\qt)} }\, e^{- \frac{\xi^2 }{T^2}} 
 e^{i \int_q \Delta \btil^a_i(q)\, \left[\phi^{1a}_i(\qp,\qt)-\phi^{2a}_i(\qp,\qt)\right]\cos(E_q\xi)} \\
 &\times&e^{-\frac{1}{4}\int_q\left\{\left[\phi^{1a}_i(q^+,\qt)-2\Delta\btil^a_i(q)\sin(E_q\xi)\right]\left[\phi^{1a}_i(q^+,-\qt)-2\Delta\btil^a_i(-q)\sin(E_q\xi)\right]+\left[\phi^{2a}_i(q^+,\qt)-2\Delta\btil^a_i(q)\sin(E_q\xi)\right]\left[\phi^{2a}_i(q^+,-\qt)-2\Delta\btil^a_i(-q)\sin(E_q\xi)\right]\right\}}\nn\\
 &=&
\frac{1}{\sqrt \pi\, T} \int d\xi\, e^{- \frac{\xi^2 }{T^2}}  e^{-\int_\qt \left\{\frac{\rho^a(\qt)\rho^a(-\qt)}{2\mu^2(\qt)} +i\Delta\btil^a_i(q)\left[\exp(iE_q\xi)\phi^{2a}_i(q)-\exp(-iE_q\xi)\phi^{1a}_i(q)\right]+2\Delta\btil^a_i(q)\Delta\btil^a_i(-q)\sin^2(E_q\xi)\right\}}\nn\\&\times&  e^{-\frac{1}{4}\int_q\left[\phi^{1a}_i(q^+,\qt)\phi^{1a}_i(q^+,-\qt)+\phi^{2a}_i(q^+,\qt)\phi^{2a}_i(q^+,-\qt)\right] } \, .
\nn
\eea
Using this expression we have
\bea
\Trace[\left(\hat\rho'_{P,\xi}\right)^N]&=&\left[ 
\frac{1}{\sqrt \pi\, T} \right]^N \int\prod_{\alpha=1}^N \left[\mathcal D \phi_i^{a\alpha} d\xi_\alpha\right]\, e^{- \frac{\xi_\alpha^2 }{T^2}}  e^{-\int_\qt \left\{2\Delta\btil^a_{i}(q)\Delta\btil^a_{i}(-q)\sin^2(E_q\xi_\alpha)\right\}}\nn\\
&\times&e^{-\int_q\left\{\frac{1}{2}\phi^{\alpha a}_i(q^+,\qt)\phi^{\alpha a}_i(q^+,-\qt)+i\left[\Delta\btil^a_{i}(q)\exp(iE_q\xi_{\alpha-1})-\Delta\btil^a_{i}(q)\exp(-iE_q\xi_{\alpha})\right]\phi^{\alpha a}_i(q)\right\}} \, .
\eea 
With the understanding $\xi_0=\xi_N$ and $\phi_0=\phi_N$, the integral over $\phi^\alpha$ is easily performed with the result
\bea
\Trace[\left(\hat\rho'^N_{P,\xi}\right)]
&=&\left[ 
\frac{1}{\sqrt \pi\, T} \right]^N \int\prod_{\alpha=1}^N \left[ d\xi_\alpha\right]\,  e^{- \frac{\xi_\alpha^2 }{T^2} }\nn\\
&\times&e^{-\int_q \left[\Delta\btil^a_{i}(q)\Delta\btil^a_{i}(-q)-\Delta\btil^a_{i}(q)  \Delta\btil^a_{i}(q)\exp\left(iE_q(\xi_{\alpha-1}-\xi_\alpha)\right)\right]} \, .
\label{rhon1}
\eea 

It is now straightforward to expand the integrand in $\Delta\btil^2$ and perform the $\xi$ integrals at any finite order. The eigenvalue $\delta_1$ then can be read off the term proportional to $N$. Keeping the first two terms we get
\beq
\delta_1=\delta_1(1)+\delta_1(2) \, , 
\eeq
with
\bea\label{delta12}
\delta_1(1)&=& \int_q \Delta \btil^2(q)\, \bigg( 1 - e^{-\frac{E_q^2T^2}{2}}  \bigg)\\
\delta_1(2)&=&\int_{q,p} \Delta \btil^2(q) \Delta \btil^2(p) \bigg( 1 - e^{-\frac{E_q^2T^2}{2}}  -e^{-\frac{E_p^2T^2}{2}}+e^{-\frac{(E_q+E_p)^2T^2}{2}}\bigg) \, .
\nonumber
\eea
Note that within our approach it is perfectly legal to keep terms in $\delta_1$ beyond the leading order in $\Delta\btil^2$. 
The first correction to $\Trace   [\hat\rho^N]$ due to $\delta_i, \ \ i\ge 2$ appears only in the order $\Delta\btil^{2N}$, 
and thus does not interfere with the extraction of higher order terms in  $\delta_1$.

Thus in the weak field limit the single event entropy,  up to logarithmic accuracy,  is given by 
\beq\label{entro}
\sigma^E_{\Delta b^2 \ll 1} \simeq - \delta_1 \log \delta_1 \approx
- \int_q \Delta \btil^2(q)\, \bigg( 1 - e^{-\frac{E_q^2T^2}{2}}  \bigg) \log\left[ \int_q \Delta \btil^2(q)\, \bigg( 1 - e^{-\frac{E_q^2T^2}{2}}  \bigg)\right]\, .
\eeq
Including the first correction of eq.(\ref{delta12}) gives
\bea\label{entro1}
\sigma^E_{\Delta b^2 \ll 1} &\approx&
- \left[ \int_q \Delta \btil^2(q)\, \bigg( 1 - e^{-\frac{E_q^2T^2}{2}}  \bigg) +\int_{q,p} \Delta \btil^2(q) \Delta \btil^2(p) \bigg( 1 - e^{-\frac{E_q^2T^2}{2}}  -e^{-\frac{E_p^2T^2}{2}}+e^{-\frac{(E_q+E_p)^2T^2}{2}}\bigg)\right]\nonumber\\
&\times&\log \left[\int_q \Delta \btil^2(q)\, \bigg( 1 - e^{-\frac{E_q^2T^2}{2}}  \bigg)\right]\, .
\eea
Note that we have not added the correction to $\delta_1$ under the logarithm, since that would exceed the leading logarithmic accuracy to which our results are valid.

\subsection{Interpreting the entropy}
Eq.(\ref{entro}) has a very simple physical interpretation. 
Recall that, within the CGC formalism, the total number  (multplicity) of soft gluons produced in a given event (at fixed $\rho^a$ and fixed $S$) is given by
\beq
 n= \int_q \Delta \btil^2(q)\, .
 \eeq
 The simplest naive expression for Bolzman  entropy of a system of $n$ particles would be
 \beq
 \sigma=-n\log n \, .
 \eeq
 Eq.(\ref{entro}) has a very similar form. If we interpret $T$ as time at which the entropy is calculated, and introduce  the number of particles produced up to time $T$ as
 \beq\label{nt}
 n(T)= \int_q \Delta \btil^2(q)\, \bigg( 1 - e^{-\frac{E_q^2T^2}{2}}  \bigg) \, , 
 \eeq
 the entropy given by eq.(\ref{entro}) becomes simply
 \beq 
 \sigma^E(T)=-n(T)\log n(T) \, .
 \eeq
Interpretation of eq.(\ref{nt}) as the number of produced particles is  also very natural. Essentially it says that up to time $T$ only those particles are actually produced, 
which have energies $E_q>1/T$. Those are precisely the particles that ``decohere'' from the rest of the outgoing wave function during the time $T$, 
since the phases of their individual  wave functions change by a number which is at least of order one. If the time $T$ is short, only very energetic gluons are produced, 
while if one waits infinite amount of time, all gluons that are present in the wave function immediately after scattering, decohere from each other and are therefore produced in the final state.

 From the point of view of our ``calorimeter'' degree of freedom this is equivalent to saying that in order to resolve produced gluons with very small energies, 
 calorimetric measurement must be allowed to take very long time $T>1/E_q$. Thus at time $T=0$ our state has vanishing entropy, 
 since it is just the pure state which emerges from the scattering region as no individual gluon states can be resolved, 
 while at $T\rightarrow \infty$ it has the entropy of a completely incoherent ensemble of all gluons produced in the given event.

\section{Time dependent entropy for the ensemble of events}\label{enenavgrho}
In the previous section we have calculated the time dependence of entropy for  single event which arises due to interaction with the calorimeter (``white noise''). 
It would be interesting to combine this with the calculation of ~\cite{Kovner:2015hga} which gave the entropy for the ensemble of events, 
but without time evolution, i.e. treating every single event density matrix as pure state. In this section we will  do just that. 

\subsection{Calculation of the entropy}
Our starting point is the expression for the soft gluon density matrix, which has to be reduced over the valence modes as well as the white noise degree of freedom. 
We use the McLerran-Venugopalan model for the distribution of the valence degrees of freedom~\cite{McLerran:1993ni,McLerran:1993ka}
\bea
 \hat{\rho}_{P,\xi,\rho}[S]
&=&\mathcal{N}
\frac{1}{\sqrt \pi\, T} \int \left[\mathcal D \rho \right]d\xi\, e^{-\int_\qt \frac{\rho^a(\qt)\rho^a(-\qt)}{2\mu^2(\qt)} }\, e^{- \frac{\xi^2 }{T^2}} 
e^{-i H_0 \xi}\, e^{i \int_q \Delta \btil^a_i(q)\, \phi^a_i(\qp,\qt)} \ketv
\brav e^{-i \int_\pt \Delta \btil^b_j(p)\, \phi^b_j(\pp,\pt)} \, e^{i H_0 \xi} \, .
\label{redrho1}
\eea
At this point our density matrix is defined for a fixed target field which enters via the dependence of $\Delta\btil$ on the Wilson line $S$.  
Since we would like to calculate the entropy for the complete ensemble of events we should also average over all possible configurations of the target. 
The consistent way of doing this is to treat the target fields as what they are, i.e.  additional quantum  degrees of freedom. 
Eq.(\ref{redrho1}) then defines the diagonal element of the density matrix in the target field basis. 
We should however reduce the density matrix over all degrees of freedom except the soft gluon fields. To achieve this we have to trace over the target fields. 
Thus we consider 
\bea
 \hat{\rho}_{P,\xi,\rho}
&=&\mathcal{N}
\frac{1}{\sqrt \pi\, T} \int \left[\mathcal D \rho \right] \left[\mathcal D S \right]d\xi\, W[S]e^{-\int_\qt \frac{\rho^a(\qt)\rho^a(-\qt)}{2\mu^2(\qt)} }\, e^{- \frac{\xi^2 }{T^2}} 
e^{-i H_0 \xi}\, e^{i \int_q \Delta \btil^a_i(q)\, \phi^a_i(\qp,\qt)} \ketv
\brav e^{-i \int_\pt \Delta \btil^b_j(p)\, \phi^b_j(\pp,\pt)} \, e^{i H_0 \xi} \, .
\nn \\
\label{redrho}
\eea
Here $W[S]$ is the normalized weight functional arising from the target wave function. We leave it unspecified for now. 

As in the previous section we now get rid of $H_0$ in the exponential using eq.(\ref{phit}).
Calculating the matrix element of the density matrix we find
\bea
\label{redmat}
&&\langle\phi^1\vert\hat\rho_{P,\xi,\rho}\vert\phi^2\rangle=
\frac{1}{\sqrt \pi\, T} \int \left[\mathcal D \rho \right]\left[\mathcal D S \right]d\xi\, W[S]\, e^{-\int_\qt \frac{\rho^a(\qt)\rho^a(-\qt)}{2\mu^2(\qt)} }\, e^{- \frac{\xi^2 }{T^2}} 
 e^{i \int_q \Delta \btil^a_i(q)\, \left[\phi^{1a}_i(\qp,\qt)-\phi^{2a}_i(\qp,\qt)\right]\cos(E_q\xi)} \\
 &\times&e^{-\frac{1}{4}\int_q\left\{\left[\phi^{1a}_i(q^+,\qt)-2\Delta\btil^a_i(q)\sin(E_q\xi)\right]\left[\phi^{1a}_i(q^+,-\qt)
 -2\Delta\btil^a_i(-q)\sin(E_q\xi)\right]+\left[\phi^{2a}_i(q^+,\qt)-2\Delta\btil^a_i(q)\sin(E_q\xi)\right]\left[\phi^{2a}_i(q^+,-\qt)-2\Delta\btil^a_i(-q)\sin(E_q\xi)\right]\right\}}\nn\\
 &=&\mathcal{N}
\frac{1}{\sqrt \pi\, T} \int \left[\mathcal D \rho \right]\left[\mathcal D S \right]d\xi\, W[S]\, e^{- \frac{\xi^2 }{T^2}}  e^{-\int_\qt \left\{\frac{\rho^a(\qt)\rho^a(-\qt)}{2\mu^2(\qt)} 
+i\Delta\btil^a_i(q)\left[\exp(iE_q\xi)\phi^{2a}_i(q)-\exp(-iE_q\xi)\phi^{1a}_i(q)\right]+2\Delta\btil^a_i(q)\Delta\btil^a_i(-q)\sin^2(E_q\xi)\right\}}
\nn\\
&\times&  
e^{-\frac{1}{4}\int_q\left[\phi^{1a}_i(q^+,\qt)\phi^{1a}_i(q^+,-\qt)+\phi^{2a}_i(q^+,\qt)\phi^{2a}_i(q^+,-\qt)\right] }\, .\nn
\eea

We will now employ the same logic as in the previous section to calculate the entropy in the weak field limit. 
We note that for weak fields, $\Delta\btil=0$, our density matrix describes a pure state. 
This means that for small $\Delta\btil$ it has one eigenvalue $\Delta=1-\delta_1$ which is close to unity, and others that are small. 
As we explained in the previous section, to calculate the entropy of such density matrix we only need to find $\delta_1$. 
We will do it, as before by calculating $\Trace [\hat\rho^N]$ for large $N$.

We will set up this calculation a little differently than in the previous section. 
We keep the reduced density matrix in the form  of eq.(\ref{redmat}) while calculating the trace. 
Obviously for each factor of $\hat\rho$ in the product of density matrices we have to introduce 
its own $\rho$ and $\xi$, and so these fields also acquire a replica index. We can then write 
\bea
\Trace[\hat\rho^N_{P,\xi,\rho}]
&=&
\left[ \mathcal{N}
\frac{1}{\sqrt \pi\, T} \right]^N \int\prod_{\alpha=1}^N \left[\mathcal D \rho_\alpha \mathcal D S_\alpha d\xi_\alpha\right]\, W[S_\alpha]e^{- \frac{\xi_\alpha^2 }{T^2}}  
e^{-\int_\qt \left\{\frac{\rho_\alpha^a(\qt)\rho_\alpha^a(-\qt)}{2\mu^2(\qt)}+2\Delta\btil^a_{i\alpha}(q)\Delta\btil^a_{i\alpha}(-q)\sin^2(E_q\xi_\alpha)\right\}}
\nn\\
&\times&e^{-\int_q\left\{\frac{1}{2}\phi^{\alpha a}_i(q^+,\qt)\phi^{\alpha a}_i(q^+,-\qt)+
i \left[\Delta\btil^a_{i(\alpha-1)}(q)\exp(iE_q\xi_{\alpha-1})-\Delta\btil^a_{i\alpha}(q)\exp(-iE_q\xi_{\alpha})\right]\phi^{\alpha a}_i(q)\right\}} \, .
\eea 
The integral over $\phi^\alpha$ is easily performed with the result
\bea
\Trace[\hat\rho^N_{P,\xi,\rho}]&=&\left[ \mathcal{N}
\frac{1}{\sqrt \pi\, T} \right]^N \int\prod_{\alpha=1}^N \left[\mathcal D \rho_\alpha \mathcal D S_\alpha d\xi_\alpha\right]\, W[S_\alpha] e^{- \left\{\int_\qt\frac{\rho_\alpha^a(\qt)\rho_\alpha^a(-\qt)}{2\mu^2(\qt)} +\frac{\xi_\alpha^2 }{T^2}\right\}}e^{-\int_q 2\Delta\btil^a_{i\alpha}(q)\Delta\btil^a_{i\alpha}(-q)\sin^2(E_q\xi_\alpha)   }\nn\\
&\times&e^{-\frac{1}{2}\int_q \left[\Delta\btil^a_{i(\alpha-1)}(q)\exp(iE_q\xi_{\alpha-1})-\Delta\btil^a_{i\alpha}(q)\exp(-iE_q\xi_{\alpha})\right] \left[\Delta\btil^a_{i(\alpha-1)}(-q)\exp(iE_q\xi_{\alpha-1})-\Delta\btil^a_{i\alpha}(-q)\exp(-iE_q\xi_{\alpha})\right]  }\nn\\
&=&\left[ \mathcal{N}
\frac{1}{\sqrt \pi\, T} \right]^N \int\prod_{\alpha=1}^N \left[\mathcal D \rho_\alpha  \mathcal D S_\alpha d\xi_\alpha\right]\,  W[S_\alpha]e^{- \left\{\int_\qt\frac{\rho_\alpha^a(\qt)\rho_\alpha^a(-\qt)}{2\mu^2(\qt)} +\frac{\xi_\alpha^2 }{T^2}\right\}}\nn\\
&\times&e^{-\int_q \left[\Delta\btil^a_{i\alpha}(q)\Delta\btil^a_{i\alpha}(-q)-\Delta\btil^a_{i(\alpha-1)}(q)  \Delta\btil^a_{i\alpha}(q)\exp\left(iE_q(\xi_{\alpha-1}-\xi_\alpha)\right)\right]} \, ,
\label{rhon}
\eea 
with the understanding $\xi_0=\xi_N$, $S_0=S_N$ and $\rho_0=\rho_N$.

This expression allows us to calculate $\delta_1$ in expansion in powers of $\mu^2$, i.e. in the weak field limit. We just expand the exponent in eq.(\ref{rhon}) to leading order in $\Delta\btil^2$. In this calculation the cross term between different replica fields vanishes due to the symmetry of the replica $\rho$ integral. Expanding and collecting terms we find
\beq
\Trace[\hat\rho^N_{P,\xi,\rho}]=1-N\langle \int_q \Delta\btil^a_{i}(q)\Delta\btil^a_{i}(-q)\rangle_{(\rho,S)}
\eeq
where we have introduced the notation $\langle \dots \rangle_{(\rho,S)}$ to denote the averages w.r.t. the projectile and the target density matrix; and thus
\beq\label{d11}
\delta_1=\langle \int_q \Delta\btil^a_{i}(q)\Delta\btil^a_{i}(-q)\rangle_{(\rho, S)}
\eeq

This is an exceedingly simple and somewhat unexpected result. 
We find that the eigenvalue and, therefore, the entropy of the density matrix averaged over the event ensemble does not depend on time in the weak field limit.
Note on the other hand that, if we do not integrate over $\rho$ (and $S$), but simply perform this calculation at fixed value of $\rho$, which we did in the previous section, 
we have an additional time dependent contribution, as discussed above.

It is also straightforward to calculate the next correction in powers of $\mu^2$. Only the terms local in replica space contribute to the result for the eigenvalue:
\beq\label{d1}
\delta_1=\langle \int_q \Delta\btil^a_{i}(q)\Delta\btil^a_{i}(-q)\rangle_{(\rho,S)}-\frac{1}{2}\left[\langle\left[\int_q \Delta\btil^a_{i}(q)\Delta\btil^a_{i}(-q)\right]^2\rangle_{(\rho,S)}+\int_{q,p}\langle 
\Delta\btil^a_{i}(q)\Delta\btil^b_{j}(p)\rangle_\rho\langle\Delta\btil^a_{i}(-q)     \Delta\btil^b_{j}(-p)\rangle_{(\rho, S)} e^{-\frac{(E_q+E_p)^2T^2}{2}}\right]
\eeq

The entropy of the whole ensemble of events is therefore 
\beq
\sigma^E=-\delta_1\log \delta_1
\eeq
with $\delta_1$ given in eq.(\ref{d1}).

We note that the time dependence does survive in eq.(\ref{d1}). Moreover its qualitative behaviour is reasonable, i.e. the absolute value of $\delta_1$  is an increasing function of  time $T$, and therefore the entropy grows with time. One does indeed expect such a trend since the entropy produced in an individual event grows with time, as soft gluons decohere from each other.

\subsection{Interpretation of the time dependence}

As noted above, the time dependence of the entropy for the ensemble of events is much weaker than for a single event. 
This begs for some explanation. We believe that the reason for this lies in the so called monogamy of entanglement~\cite{Koashi:2004men}. 
It is known that if a system $A$ is maximally entangled with system $B$, then none of those systems can be entangled with the third system $C$. 
A plausible interpretation of time independence of eq.(\ref{d11})  is that for very weak fields the soft gluon degrees of freedom are maximally entangled with the valence charges. 
By the monogamy of entanglement it then means that coupling to another degree of freedom (in our case $\xi$) does not change the entropy of $\phi$. 
For stronger fields the entanglement is not maximal and some time dependence survives, but it is significantly weaker than for a single event.

\section{Conclusions}\label{conclusions}

In this paper, we have extended the approach of~\cite{Kovner:2015hga}, by considering the entropy 
of the soft gluons system produced in a single event in  high-energy p-A collisions.  Strictly speaking, the system 
of soft gluons produced in a single event is in a pure state and has zero entropy. 
However in the energy eigenstate basis the off diagonal matrix elements of the density matrix are strongly oscillating functions of time. 
When averaged over time they vanish, and such density matrix corresponds to a mixed state with finite von Neuman entropy.

Physically averaging over time is always performed since any experimental apparatus has finite time and energy resolution. 
We therefore define the time dependent entropy as entanglement entropy of the system of soft gluons and the ``calorimeter'' - 
an auxiliary degree of freedom representing an apparatus that measures the energy of the outgoing gluons.

We calculate the time dependent  single event entropy in the limit of weak fields
and observe that it has a simple interpretation as the Bolzman entropy of the system of gluons produced up to the time at which the entropy is calculated.

We also calculate the time dependent entropy for the ensemble of events, and find that it has a much weaker time dependence than the analogous quantity for an individual event. 
On a qualitative level we attribute this feature to the monogamy of entanglement.

Our investigation in this paper is of an academic nature given the fact that entropy is not a directly measurable quantity. 
We hope however that this will be an important step towards understanding of quantum dynamics of colliding systems. 
In particular time evolution of entanglement observed in atomic systems has been related to possible eigenstate thermalization~\cite{Srednicki:1994eth},
of which some tentative evidence has recently been reported in condensed matter experiments~\cite{Kaufman:2016}. 
Analogous phenomenon could arise in high energy scattering potentially explaining important features of observed spectra of produced particles~\cite{Baker:2017wtt,Kharzeev:2017qzs}. 
Momentum space entanglement considered in this paper is {\it a priori} as good a source of entropy and disorder, possibly leading to thermalization, as its coordinate space counterpart. 
Since the CGC wave function is an explicit and controlled model of hadronic wave function, 
it provides an interesting testing ground for this type of ideas. We hope to investigate this aspect of entanglement in future work.

\appendix

\section{The coherent operator and the Heisenberg evolution operator}\label{coherent}

In this Appendix we clarify the relation between the coherent operator $\Omega$, which is used in this paper to construct the CGC state in perturbation theory,
and the Heisenberg evolution operator of the system of soft gluons.

Recall that $\Omega$ is defined as the operator that diagonalizes the QCD light cone Hamiltonian of the soft modes in the background of valence sources. 
The LO QCD Hamiltonian in the light-cone gauge ($A^+ = 0$) with which we work is given 
by a free part $H_0$ plus an $\mathcal{O}(g)$ correction, $H_g$
\bea
H_{LC}
&=& 
H_0 + H_g \, ,
\nn \\
H_0 
&=& 
\int_0^\infty \frac{dq^+}{2\pi} \int\frac{d^2\qt}{(2\pi)^2} \frac{\qt^2}{2\qp}\, a^{\dagger a}_i(\qp,\qt) a^a_i(\qp,\qt) \equiv
\int_0^\infty \frac{dq^+}{2\pi} \int\frac{d^2\qt}{(2\pi)^2} E_q\, a^{\dagger a}_i(\qp,\qt) a^a_i(\qp,\qt) 
\nn \\
H_g
&=& 
\int_q \frac{g \qt_i \rho^a(\qt)}{\sqrt{2} \vert q^+ \vert^{3/2}} \left[  a^a_i(\qp,\qt) - a^{\dagger a }_i(\qp,-\qt) \right] \, .
\label{hamiltonian}
\eea
%


%
%
It is straightforward to show that at the leading order in the coupling constant, the vacuum state wave function of the soft gluons  is given by the action of the coherent operator $\Omega$
on the soft space Fock vacuum~\cite{Kovner:2005jc}, which we denote by $\vert 0 \rangle$,
\beq
\vert s \rangle = \Omega\, \ketv \, , 
\quad
\Omega = \exp\bigg\{i \int_q \btil^a_i(q) \, \left[ a^a_i(\qp,\qt)  + a^{\dagger a}_i(\qp,-\qt)  \right] \bigg\} 
\equiv
\exp\left\{i\,\omega \right\} \, .
\eeq
The physical interpretation of $\Omega$
is to dress the soft vacuum with the Weizs\"acker-Williams  transverse field generated  by the boosted valence color charges.

The canonical commutation relations lead to
\beq
\Omega^\dagger a^{\dagger a}(q)\Omega=-i\btil^a_i(\qp,\qt); \ \ \ \ \ \Omega^\dagger a^{a}(q)\Omega=i\btil^a_i(\qp,-\qt)\, .
\eeq  
Therefore the coherent operator $\Omega$ unitarily transforms the LO QCD light-cone Hamiltonian into diagonal form
\beq
\Omega H_{LC} \Omega^{\dagger} = \Omega \left( H_0  +H_g \right) \Omega^{\dagger} =H_0 - \int_q \frac{g^2 \rho^a(\qt) \rho^a(-\qt)}{(\qp)^2} \equiv H_0 + F(\rho)\, .
\label{OmegaH}
\eeq
To this order the spectrum of the theory is identical to the free spectrum, and all the eigenstates are just unitarily transformed Fock space states.
 Thus an arbitrary eigenstate has the form
 \beq\label{intf}
 \overline{\vert p_1...p_n\rangle}=\Omega\vert p_1...p_n\rangle \, .
 \eeq
 where $\vert p_1...p_n\rangle$ is he $n$-gluon Fock state. This follows since (we neglect the constant zero point energy) the Fock states are eigenstates of the free Hamiltonian, and thus
 \beq
 H\overline{\vert p_1...p_n\rangle}=\Omega\Omega^\dagger H\Omega\vert p_1...p_n\rangle=\Omega H_0\vert p_1...p_n\rangle=\Big(E(p_1)+...E(p_n)\Big)\Omega \vert p_1...p_n\rangle=\Big(E(p_1)+...E(p_n)\Big)\overline{\vert p_1...p_n\rangle} \, .
 \eeq

 Now recall that the Heisenberg evolution operator for evolution over the infinite time interval is defined as
 \beq
 U(-\infty,0)=\exp\{-i\int_{-\infty}^0H(t) dt\} \, .
 \eeq
Although formally the Hamiltonian is time independent, to develop perturbation theory, and in general a well defined  path integral representation, 
one regulates it so that the interaction vanishes at time infinitely far  from the interaction region, i.e. $H(t\rightarrow -\infty)=H_0$.
 
 The regulated evolution operator has a very important property, namely that if one starts at $t\rightarrow -\infty$ from an eigenstate of the free Hamiltonian, under the action of $U(-\infty,0)$ it evolves to an eigenstate of the interacting Hamiltonian . This can be written
 as
 \beq \label{intf1}
 U(-\infty,0)\vert E_0\rangle_0= e^{i\Phi}\overline{\vert E\rangle} \, , 
\eeq
 where the state $\vert E_0\rangle_0$ is the eigenstate of the noninteracting Hamiltonian with eigenvalue $E_0$, the state $\overline{\vert E\rangle}$ is the eigenstate of the full Hamiltonian with eigenvalue $E$, and $\Phi$ is an arbitrary phase, which in principle can be defined to be zero.
 
 Comparing Eqs. (\ref{intf}) and (\ref{intf1}) we see that they are identical. We therefore conclude that
 \beq
 \Omega=U(-\infty,0)
 \eeq
 and by the same token
  \beq
 \Omega^\dagger=U(0,\infty) \, .
 \eeq
 This concludes our demonstration.
 

\section*{Acknowledgments}
AK thanks the Ben Gurion University of the Negev for hospitality during the period when this project was started.
AK was supported by the NSF Nuclear Theory grant 1614640, the Fulbright US scholar program and  CERN scientific associateship.
MS and ML are grateful to the CERN theory division, particularly to its Heavy Ion group, 
for support and hospitality at the time when this work was being completed. 
MS is supported by the  Israeli Science Foundation through grant 1635/16, 
by the BSF grants 2012124 and 2014707, by the COST Action CA15213 THOR
and by a Kreitman fellowship by the Ben Gurion University of the Negev.

\end{document}